\documentclass[12pt]{article}

\usepackage{amsfonts,amssymb}
\newcommand{\dd}{ \hbox{d} }

\begin{document}

\title{\bf An Application of \\
the Schwarzian Derivative\thanks{Talk given by B. Zhou at the 
CCAST Workshop on
Integrable System, Beijing, May 4--7. }}
\author{Bihn Zhou\thanks{E-mail: zhoub@itp.ac.cn}\, and 
Chuan-Jie 
Zhu\thanks{
E-mail: zhucj@itp.ac.cn} \\
Institute of Theoretical Physics, Chinese Academy of\\
Sciences, P. O. Box 2735, Beijing 100080, P. R. China}

\maketitle

\begin{abstract}
In this note we present an application of the Schwarzian derivative. By
exploiting some properties of the Schwarzian derivative, we solve the
equation appearing in the gravity-dilaton-antisymmetric tensor system.
We also mention  that this method can also be used to solve
some other equations.

\end{abstract}

The Schwarzian derivative appears naturally in the transformation law of the
two-dimensional stress-energy tensor \cite{Joe}:
\begin{equation}
(\partial_z z')^2 \, T'(z') = T(z) - { c \over 12} \, \{z', z\}, 
\end{equation}
where $\{z', z\}$ denotes the Schwarzian derivative:
\begin{equation}
S(g) \equiv  \left( \frac{g''}{g'} \right)' - \frac{1}{2} \left( \frac{g''}{g'}
  \right)^2.
\label{Schw}
\end{equation}
It comes as a surprise that it also appears in a differential
equation which we obtained during the course of 
solving the equations of motion of the coupled system of
gravity-dilaton-antisymmetric tensor system.  The last equation to be
solved \cite{ZhouZhua,ZhouZhub} is:
\begin{equation}
Y^{\prime\prime} - \frac{\Delta}{2}\, (Y^\prime)^2 + Q(r)Y^\prime = R(r),
\label{NDf}
\end{equation}
where $Q(r)$ and $R(r)$ are two specific functions of $r$ (see \cite{ZhouZhub}
for exact formulas) and $\Delta$ is a (non-vanishing) constant.

To solve this equation completely, we follows 
\cite{ZhouZhua, ZhouZhub}. Let
\begin{equation}
g = \int \dd \, r \, e^{ \Delta Y - \int Q(r) dr },
\end{equation}
or,
\begin{equation}
 Y =  \frac{1}{\Delta}\left( \ln (g') + \int Q(r)dr \right),
\label{f2Y}
\end{equation}
eq. (\ref{NDf}) becomes
\begin{equation}
\left( \frac{g''}{g'} \right)' - { 1 \over 2 } \, \left( \frac{g''}{g'}
\right)^2 = \tilde{R}(r).
\label{lasty}
\end{equation}
Here
\begin{eqnarray}
  \tilde{R}(r) & = & \Delta R(r) - Q^\prime(r) - \frac{1}{2}\, Q^2(r)
  \nonumber \\
    & = & -\frac{\tilde{d}^2-1}{2\, r^2} +
    \frac{\tilde{\Lambda}\, r^{2\tilde{d}-2}}{ 2\,(C_2+C_1 r^{2\tilde{d}})^2 }
\label{NRdef}
\end{eqnarray}
with
\begin{equation}
  \tilde{\Lambda} =\Delta \, K - 4\, \tilde{d}^2\, C_1\, C_2
\end{equation}
a constant.

The left-hand side of eq. (\ref{lasty}) is the Schwarzian derivative of the
function $g$ as we defined in 
(\ref{Schw}). Thus, in order to solve eq. (\ref{lasty}), one must find a
function $g$ such that
\begin{equation}
S(g)= -\frac{\tilde{d}^2-1}{2r^2} + \frac{\tilde{\Lambda}}{2}\
  \frac{ r^{ 2\tilde{d}-2 } }{ (C_2 + C_1 r^{2\tilde{d}})^2 }.
\label{target}
\end{equation} 
We will do this in the following sections \cite{ZhouZhua,ZhouZhub}.

\section{Notations and Conventions}

Let $f$ be an arbitrary function of $r$ and $f'$ be its derivative. The notation
$f(r)$ always represents the value of the function $f$ at $r$. Similarly, 
$f'(r)$ is the value of the function $f'$ at $r$.

For two functions $f$ and $g$, $f \,g$ is their product\footnote{We
denote $f\,f$ by $f^2$, $f\, f\, f$ by $f^3$, and so on.},
and $\displaystyle{\frac{f}{g}}$ is their division, while
the composition of two functions $f$ and $g$ is denoted by $f\circ g$.
These functions are defined by
\begin{eqnarray}
   (f\, g)(r)  & = & f(r) g(r), \\
  \left( \frac{f}{g} \right)(r) & = &  \frac{ f(r) }{ g(r) }, \\
  (f\circ g)(r) & = &  f( g(r) ).
\end{eqnarray}
For product of three or more functions we assume that the composition of two 
functions
has a higher rank of precedence of associavity than the products of functions.
Nevertheless the symbol
$\displaystyle{\frac{f}{g}}$ is considered to be a pure entity and can't be
broken. To understand these conventions we have the following examples:
\begin{eqnarray}
  && f\, g\circ h =  g\circ h\, f = f\, (g\circ h), \\
  && \frac{f}{g}\circ h = \left( \frac{f}{g} \right)\circ h
  = \frac{f\circ h}{g\circ h}.
\end{eqnarray}

If $k\in \mathbb{C}$ or $\mathbb{R}$, we  define a function $l_k$ as
the multiplication of its variable with the number $k$, i.e.,
$$
  l_k(r)=k\, r.
$$
For convenience we also consider $k$ itself as a constant function taking the
value  $k$:
$$
  k(r) = k.
$$
Other functions such as the power function $r^s$ with real number $s$
and the exponential function $e^r$ are denoted by $p_s$ and $\exp$ and 
other elementary functions are denoted by their standard mathematical 
symbols.

With these conventions, we can derive the following derivative rules:
\begin{equation}
\begin{array}{ll}
  l'_k=k, &  \qquad \qquad k'=0, \\
  p'_s=s\, p_{s-1}  & \qquad \qquad   \arctan'=\frac{1}{1+p_2}, \\
  \exp'=\exp, & \qquad \qquad  \ln'=p_{-1},   
\end{array}
\end{equation}
and the derivative rule for composition of function is
\begin{equation}
  (f\circ g)'=g'\ f'\circ g
\end{equation}

\section{Some Properties of the Schwarzian Derivative}

 Now we list some elementary properties of the Schwarzian derivative.

(1) If $f$ and $g$ are two functions, we have 
\begin{equation}
  S(f\circ g)=(g')^2\ S(f)\circ g + S(g).
\end{equation}

(2) If $f=\displaystyle{ \frac{l_a+b}{l_c+d}}$ for some numbers $a$, $b$, $c$
and $d$, namely, $f(r)=\displaystyle{ \frac{ar+b}{cr+d} }$, then
\begin{equation}
  S\left(\frac{l_a+b}{l_c+d}\right)=0.
\end{equation}
This is the well-know fact that the fractional linear transformation is a
global conformal transformation of the complex sphere.  By using this result
we have
\begin{equation}
  S\left( \frac{l_a+b}{l_c+d}\circ g \right) = S(g).
\end{equation}
That is to say, if $g$ is a special solution of the equation $S(f)=R$, 
the general solution will be
\begin{equation}
  f = \frac{l_a+b}{l_c +d}\circ g = \frac{ag+b}{cg+d}
\label{fraclinear}
\end{equation}
with constants $a$, $b$, $c$, $d$ such that $ad-bc=1$.

(3) For $s\in\mathbb{R}$,  we have 
\begin{eqnarray}
 S(l_k)  & =&  0, \\
 S(p_s) & =&  -\frac{s^2-1}{2p_2}, \\
 S(\exp) & =&  - \frac{1}{2}, \\
 S(\ln) & =&  \frac{1}{2p_2}, \\
 S(\tan) & =&  2, \\
S(\arctan) & =&  - \frac{2}{(1+p_2)^2}.
\end{eqnarray}

\section{The Complete Solution of Eq. (\ref{target})}

The function $\tilde{R}$ in Eq. (\ref{target}) is
\begin{equation}
  \tilde{R} = -\frac{\tilde{d}^2-1}{2p_2}
  + \frac{ \tilde{\Lambda} p_{2\tilde{d}-2} }{2}
  \,  \frac{1}{(C_2+C_1\, p_2)^2}\circ p_{\tilde{d}}.
\end{equation}
For $C_1 C_2 > 0$, we have 
\begin{eqnarray}
  \tilde{R} &=& S(p_{\tilde{d}}) + (p'_{\tilde{d}})^2\ \, 
S(h_1)\circ p_{\tilde{d}}
\nonumber \\
  &=& S(h_1\circ p_{\tilde{d}})
\label{Sh1}
\end{eqnarray}
where $h_1$ is a yet-unknown function such that
\begin{eqnarray}
  S(h_1) &=& \frac{\tilde{\Lambda}}{2\tilde{d}^2C_2^2}\
  \frac{1}{(1+\frac{C_1}{C_2}p_2)^2}
\nonumber \\
  &=& S(l_{\sqrt{C_1/C_2}}) + (l'_{\sqrt{C_1/C_2}})^2\ \frac{\tilde{\Lambda
}}
  {2\tilde{d}^2 C_1 C_2}\ \frac{1}{(1+p_2)^2}\circ l_{\sqrt{C_1/C_2}}
\nonumber \\
  &=& S(h_2\circ l_{\sqrt{C_1/C_2}}).
\label{Sh2}
\end{eqnarray}
Now we want to find a function $h_2$ such that
\begin{eqnarray}
  S(h_2) &=& \frac{\tilde{\Lambda}}{2\tilde{d}^2 C_1 C_2}\ \frac{1}{(1+
p_2)^2}
\nonumber \\
  &=& S(\arctan) + \frac{ \Delta \, K}{2\tilde{d}^2 C_1 C_2} \frac{1}{(1+
p_2)^2}
\nonumber \\
  &=& S(\arctan) + (\arctan')^2 \frac{\Delta \, K}{2\tilde{d}^2 C_1 C_2}\circ
 \arctan
\nonumber \\
  &=& S(h_3\circ \arctan).
\end{eqnarray}
Note that in the third line we have considered $\displaystyle{\frac{\Delta \, K}
{2\,\tilde{d}^2 \,C_1\, C_2}}$ to be a constant function. The function $h_3$ 
in the
above is easy to find to be $\tan\circ l_k$ with
\begin{equation}
  k = \frac{1}{2\tilde{d}}\sqrt{ \frac{\Delta \, K}{C_1 C_2} },
\end{equation}
because
\begin{eqnarray}
  S(h_3) &=& \frac{\Delta \, K}{2\tilde{d}^2 C_1 C_2}
\nonumber \\
  &=& S(l_k) + (l'_k)^2\ S(\tan)\circ l_k
\nonumber \\
  &=& S(\tan\circ l_k).
\end{eqnarray}

Combining all the above steps, one finds that
\begin{equation}
  \tilde{R}=S(\tan\circ l_k\circ\arctan\circ l_{\sqrt{C_1/C_2}}\circ
  p_{\tilde{d}}).
\end{equation}
and a special solution of eq. (\ref{target}) is found to be:
\begin{equation}
  g_0=\tan\circ l_k\circ\arctan\circ l_{\sqrt{C_1/C_2}}\circ p_{\tilde{d}},
\end{equation}
namely,
\begin{equation}
  g_0(r)=\tan\left( k\arctan \sqrt{\frac{C_1}{C_2}}r^{\tilde{d}} \right).
\label{special}
\end{equation}
The special case considered in  \cite{ZhouZhuc} corresponds to $k =1$.
By using Mathematica, one can easily check that the above function is indeed
a solution of eq. (\ref{target}).

The general solution of eq. (\ref{target}) is obtained from the above special 
solution,
eq. (\ref{special}), by an  arbitrary $SL(2,R)$ transformation:
\begin{equation}
g(r) = { a_0 \, g_0(r) + b_0 \over c_0 \, g_0(r) + d_0 },
\end{equation}
where $a_0$, $b_0$, $c_0$ and $d_0$ consist of an $SL(2,R)$ matrix:
\begin{equation}
a_0 \, d_0 - b_0 \, c_0 = 1.
\end{equation}
Here we have three independent constants. This is the
right number for a third order ordinary differential equation. So we obtain the
complete solution to eq. (\ref{target}). 

With this general solution in hand one can proceed to obtain the general
solution of (\ref{NDf}). For details please see \cite{ZhouZhub}.

\section{More examples}

There are also other examples which can  be solved by the above method.
One example comes from 
the special case  $\tilde{d}=0$ and the equation is as follows ($\Delta=a^2$):
\begin{equation}
  Y'' - \frac{a^2}{2}(Y')^2 + Q(r)\ Y' = R(r)
\end{equation}
where
\begin{eqnarray}
  Q(r) &=& \frac{1+a^2}{r} + \frac{1+aC_3}{r \ln \frac{r_0}{r}}, \\
  R(r) &=& \frac{a^2}{2r^2} + \frac{1+aC_3}{r^2\ln \frac{r_0}{r}}
	+ \frac{C_3^2}{2r^2\left(\ln\frac{r_0}{r}\right)^2}
	+ \frac{(D-3)C_4^2-4C_4-4(D-1)}{4(D-2)r^2
	\left(\ln\frac{r_0}{r}\right)^2}.
\end{eqnarray}
In terms of
\begin{equation}
  g(r) = \int dr e^{a^2 Y - \int Q(r)dr},
\end{equation}
or, equivalently,
\begin{equation}
  Y = \frac{1}{a^2}\ (\ln(g') + \int Q(r)dr ),
\end{equation}
the equation for $Y$ can be written as
\begin{equation}
  S(g) = \tilde{R}(r),
\end{equation}
where $S(g)$ is the Schwarzian derivative of $g$ and
\begin{eqnarray}
  \tilde{R}(r) &=& a^2\ R(r) - Q'(r) - \frac{1}{2}\ Q^2(r) \nonumber \\
  &=& \frac{1}{2r^2} + \frac{1-K}{2r^2\left(\ln\frac{r_0}{r}\right)^2}
\end{eqnarray}
with
\begin{equation}
  K = 4(1+aC_3)+
\frac{4(D-1)+4C_4-(D-3)C_4^2}{2(D-2)}\, a^2.
\end{equation}
For different choices of $K$ the special solution is as follows:

(1) For $K=0$,  \begin{equation}
  g_0 = \ln\left|\ln\frac{r_0}{r}\right|;
\end{equation}

(2) For $K>0$ ($k=\sqrt{K}$),  
\begin{equation}
  g_0 = \left|\ln\frac{r_0}{r}\right|^k;
\end{equation}

(3) For $K<0$ ($k=\sqrt{-K}$),
\begin{equation}
  g_0 = \tan\left(\frac{k}{2}\ln\left|\ln\frac{r_0}{r}\right|\right).
\end{equation}

For more examples and extensive discussions,  we refer the readers
to \cite{Zhou}.

\section *{Acknowledgments}
We would like to thank Han-Ying Guo, Yi-hong Gao, Ke Wu, Ming Yu,
Zhu-jun Zheng and Zhong-Yuan Zhu for discussions. This work is supported in
part by funds from Chinese National Science Foundation and Pandeng Project.
C.-J. Zhu would like to thank the Abdus Salam International Center for
Theoretical Physics for hospitality at the Extended Workshop on String Theory
(June 1st--July 16th, 1999, Trieste, Italy). He would also like to thank
SISSA/ISAS (International School for Advanced Studies, Trieste, 
Italy) for hospitality where we 
finally find the time to finish the writing of  this note.


\begin{thebibliography}{[20]}

\bibitem{Joe}
 J. Polchinski, {\it String Theory},  2 vols., Cambridge
University Press, 1998.

\bibitem{ZhouZhua} B. Zhou and C. -J. Zhu, {\it The Complete Brane
Solution in 
D-Dimensional Coupled Gravity System}, preprint hep-th/9904157.

\bibitem{ZhouZhub} B. Zhou and C. -J. Zhu, {\it The Complete Black 
Brane Solution in
D-Dimensional Coupled Gravity System}, preprint hep-th/9905146.

\bibitem{ZhouZhuc} B. Zhou and C. -J. Zhu, {\it A Study of Black Brane 
Solutions in
D-Dimensional Coupled Gravity System}, preprint hep-th/9903118.


\bibitem{Zhou} B. Zhou, ITP Doctoral Thesis, 1999.

\end{thebibliography}
\end{document}